\documentstyle{article}
\begin{document} 
\newcommand{\beq}{\begin{equation}}
\newcommand{\eeq}{\end{equation}} 
\newcommand{\bea}{\begin{eqnarray}}
\newcommand{\eea}{\end{eqnarray}} 
\newcommand{\nn}{\nonumber}
\newcommand{\bal}{\begin{array}{ll}} 
\newcommand{\eal}{\end{array}}
\newcommand{\la}[1]{\lambda^{#1}}
\def\1{{\rm 1 \kern -.10cm I \kern .14cm}} \def\R{{\rm R \kern -.28cm I
\kern .19cm}}

\begin{titlepage} 
\begin{flushright}  LPTHE-ORSAY 96/35 \\hep-ph/96yymmnn
\end{flushright} 
\vskip .8cm 
\centerline{\LARGE{\bf {Neutrino masses and family symmetries}}
\footnote{To be published in Proceedings of the XXXIst Rencontres de
Moriond on "Electroweak Interactions and Unified Theories", Les Arcs,
Savoie, France, March 16-23, 1996.}}
\vskip 1.5cm \centerline{\bf {St\'ephane Lavignac
\footnote{Supported in part by the CEC SCIENCE PROJECT SC1-CT91-0729.}}}  
\vskip .5cm   \centerline{\em
Laboratoire de Physique Th\'eorique et Hautes Energies
\footnote{Laboratoire associ\'e au CNRS-URA-D0063.}} \centerline{\em
 Universit\'e Paris-Sud, B\^at.
211,} \centerline{\em F-91405 Orsay Cedex, France }

\vskip 2cm \centerline{\bf {Abstract}}

\indent
The hypothesis of an additional abelian symmetry acting in a different
way on the three families of leptons leads to interesting predictions in
the neutrino sector. Contrary to what happens in most seesaw models, the
structures of the Dirac and the Majorana matrices are determined without
any ansatz, and the neutrino masses and mixing angles are fixed once the
lepton charges under the family symmetry have been chosen. Two explicit
models using this idea are presented.

\end{titlepage} 

%%%%%%%%%%%%%%%%%%%%%%%%%%%%%%%%%%%%%%%%%%%%%%%%%%%%%%%%%%%%%

\section{Introduction}

 \indent
The problem of whether the neutrinos are massive or not is fundamental
both for theoretical and phenomenological reasons.

From a theoretical point of view, despite the absence of any
experimental evidence for nonzero neutrino masses (the present upper
bounds are $m_{\nu_{e}} < 5.1 \; eV$, $m_{\nu_{\mu}} < 160 \; keV$,
$m_{\nu_{\tau}} < 24 \; MeV$), there is no reason to expect the
neutrinos to be massless. Indeed, while the photon mass is protected by
the $ U(1)_{\rm em} $ gauge symmetry, neutrino masses are not forbidden
by any fundamental symmetry. They are only protected by lepton number
symmetry, which is an accidental global symmetry of the Standard
Model\footnote{Moreover, lepton number violation occurs in numerous
extensions of the Standard Model.}. Now, if the neutrinos are massive,
the rather unnatural suppression of their masses relative to the quarks
and charged leptons of the same family has to be explained.

On the phenomenological side, massive neutrinos can oscillate from one
flavour to another, and this phenomenon could account for the
experimental data on solar and atmospheric neutrinos, as well as the
recent LSND results. Furthermore, a neutrino with mass in the $1-10 \;
eV$ range would be a good candidate for hot dark matter, and could solve
several cosmological problems, such as structure formation or cosmic
microwave background anisotropies.

%%%%%%%%%%%%%%%%%%%%%%%%%%%%%%%%%%%%%%%%%%%%%%%%%%%%%%%%%%%%%

\section{Models of neutrino masses}

\subsection{Generalities}

 \indent
Various models have been proposed for neutrino mass. All of them need an
extension of the particle content of the Standard Model. A Dirac mass
term (${\cal L}_m = - \, m \, \bar \nu_L \nu_R + h.c.$) requires the
existence of a right-handed ($RH$) neutrino $\nu_R$ in addition to the
standard left-handed ($LH$) neutrino $\nu_L$. A tree-level Majorana mass
term (${\cal L}_m = - 1/2 \, m \, \bar \nu_L \nu^c_R + h.c.$) involves a
transition from a $I_W = - 1/2$ state ($\nu_R^c$) to a $I_W = + 1/2$
state ($\nu_L$), and must therefore originate from a Yukawa coupling to
a weak Higgs triplet (Gelmini-Roncadelli model). Other models appeal to
a particular mechanism to generate a neutrino mass. In the seesaw
mechanism \cite{seesaw}, a small Majorana mass for the standard neutrino
is induced from heavy $RH$ neutrino exchange. In charged Higgs models, a
small Majorana mass is generated from loop diagrams involving charged
Higgs bosons.

Among these models, the most popular one is the seesaw mechanism,
because it naturally leads to a very small neutrino mass. Let us
illustrate this in the one-family case. The particle content of The
Standard Model is extended to include, in addition to the ordinary $LH$
neutrino $\nu_L$, a $RH$ neutrino $N_R$. Such Standard Model singlets
are present in numerous extensions of the Standard Model (like $SO(10)$
GUT's or string models). The general neutrino mass term one considers
has the following form:
\begin{equation}
  - {1 \over 2} \left( \bar \nu_L \bar N_L^c \right)
  \left( \begin{array}{cc}
	  	0  &  m \\
		m  &  M
	 \end{array}  \right)
  \left( \begin{array}{c}
		\nu_R^c  \\  N_R
	 \end{array}  \right) + h.c.
\end{equation}
The $\Delta I_W = 1$ entry of the mass matrix is zero because one
assumes that there is no Higgs triplet, hence it is impossible to write
a Majorana mass term for $\nu_L$. The $\Delta I_W = 1/2$ entry is
protected by the electroweak symmetry, so the Dirac mass m is expected
to be of the order of the breaking scale $M_{weak} = 246 \; GeV$. On the
contrary, the $\Delta I_W = 0$ entry is not protected by any symmetry,
therefore the Majorana mass for $N_R$ can be very large (typically $M
\sim 10^{13} - 10^{14} \; GeV$ in realistic seesaw models). With this
hierarchy between the mass matrix entries, the diagonalization leads to
a hierarchical mass spectrum:
\begin{eqnarray}
  m_1 \simeq \frac{\; m^2}{M}  &  &  m_2 \simeq M
\end{eqnarray}
and a small mixing angle between the two mass eigenstates:
\begin{equation}
  \tan \theta \simeq \frac{m}{M}
\end{equation}
Thus, we end up in a natural way with a very light neutrino, with a mass
$m_1$ far below the weak scale, and a heavy neutrino. Since the mixing
angle is small, the light neutrino is mainly the standard $\nu_L$.

It is interesting to note that the presence of a zero in the (1,1)
entry, which is due to a gauge symmetry, provides us with a relation
between the mass eigenvalues and the mixing angle. Diagonalizing the
mass matrix:
\begin{displaymath}
  \left( \begin{array}{cc}
	  	0  &  m \\
		m  &  M
	 \end{array}  \right)  =
  \left( \begin{array}{rr}
	  	  \cos \theta  &  \sin \theta \\
		- \sin \theta  &  \cos \theta
	 \end{array}  \right)
  \left( \begin{array}{cc}
	  	- m_1  &   0 \\
		   0   &  m_2
	 \end{array}  \right)
  \left( \begin{array}{rr}
	  	\cos \theta  &  - \sin \theta \\
		\sin \theta  &    \cos \theta
	 \end{array}  \right)
\end{displaymath}
and writing that the (1,1) entry is zero:
\begin{displaymath}
  0 = - m_1 \cos^2 \theta + m_2 \sin^2 \theta
\end{displaymath}
we obtain the mass-angle relation
\begin{equation}
  \tan \theta = \sqrt{\frac{m_1}{m_2}}
\label{eq:angle1}
\end{equation}
This suggests that symmetries may play a crucial role in constraining
the neutrino mass and mixing pattern.

\subsection{Explicit seesaw models}

 \indent
Let us now see how the seesaw mechanism is implemented in usual models.
With one right-handed neutrino per family, the neutrino mass matrix is a
6x6 matrix:
\begin{equation}
  \left( \begin{array}{cc}
		0	&  {\cal M}_D \\
	{\cal M}^T_D    &  {\cal M}_M
	 \end{array}  \right)
\end{equation}
The masses and mixing angles of the light eigenstates are obtained from
the diagonalization of the light neutrino mass matrix
\begin{eqnarray}
 {\cal M}_{\nu} \; \; \; & = & \; \; \; \; \; \; \; \;
		- {\cal M}_D {\cal M}^{-1}_M {\cal M}^T_D  \nonumber \\
		\nonumber \\
		& = & R_{\nu}  \left( \begin{array}{ccc}
			m_{\nu_{e}} & 0 & 0 \\
			0 & m_{\nu_\tau} & 0 \\
			0 & 0 & m_{\nu_{\tau}}
			\end{array} \right)  R^T_{\nu}
\end{eqnarray}
Note that the mixing angles relevant for neutrino oscillations are given
by the analog of the CKM matrix, which also involves the charged lepton
sector diagonalization matrix $R_{e}$:
\begin{equation}
  V_L = R_{\nu} R^{\dagger}_{e}
\end{equation}

The natural scale of the Dirac matrix ${\cal M}_D$ is $M_{weak}$,
whereas the entries of the Majorana matrix ${\cal M}_M$, being not
constrained by any symmetry, are expected to be much larger than
$M_{weak}$. Apart from these restrictions, the entries of both the Dirac
and the Majorana matrices are free parameters, and one has to make a
specific ansatz in order to constrain the neutrino mass spectrum.

It is usually assumed that the Dirac mass matrix has the same structure
than the up quark mass matrix\footnote{This arises naturally in Standard
Model extensions with a quark/lepton symmetry, like the $SO(10)$ GUT.}:
${\cal M}_D \sim {\cal M}_U$. For the Majorana matrix, however, no such
simplifying assumption can be done, and it is necessary to choose a
specific structure. Various ans\"{a}tze for ${\cal M}_M$ have been
studied in the literature: degenerate (all eigenvalues are equal),
hierarchical, democratic (all entries are 1). It follows that the
neutrino spectrum of a given model depends on the ansatz that has been
chosen, which is not very satisfactory.

This problem can be evaded if one assumes that the structures of the
Dirac and the Majorana matrices are determined by a symmetry. This
symmetry has to act in a different way\footnote{Unless one envisages the
case of almost degenerate neutrinos.} on the three neutrino families,
otherwise the matrices would be unconstrained. Such a symmetry is called
a {\it family symmetry}. This approach has proven to be successful in
the quark sector, where, following the original idea by Froggatt and
Nielsen \cite{FN}, several groups \cite{LNS,IR,BR,JS,DPS} have shown
that an abelian family symmetry can reproduce the observed mass and
mixing hierarchy.

%%%%%%%%%%%%%%%%%%%%%%%%%%%%%%%%%%%%%%%%%%%%%%%%%%%%%%%%%%%%%%%%%%%%%%

\section{Fermion masses and family symmmetry}

\subsection{Quark sector}

 \indent
First of all, let us stress the motivations for introducing a new
symmetry. The experimental data show a strong hierarchy between the
fermion masses, e.g. $m_u \ll m_c \ll m_t$ in the up quark sector. The
up quark mass matrix can then be written, in first approximation:
\begin{eqnarray}
  \frac{{\cal M}_u}{m_t}  &  \simeq  &  \left( \begin{array}{ccc}
	0  &  0  &  0 \\
	0  &  0  &  0 \\
	0  &  0  &  1   \end{array}  \right)
  \label{eq:hierarchy}
\end{eqnarray}
with small corrections of order ${\cal O} \left( m_u / m_t \right)$ and
${\cal O} \left( m_c / m_t \right)$. Now the zero entries can be
interpreted as zeroes induced by a symmetry (the corresponding Yukawa
couplings being forbidden by the symmetry), and the small corrections as
arising through the breaking of this symmetry.

Let us now see how this scenario can be realized with an abelian family
symmetry. We extend the gauge group of the Minimal Supersymmetric
Standard Model (MSSM) with a family symmetry $U(1)_X$ ($X$ denotes the
conserved charge associated with the symmetry). Each Yukawa coupling
$Q_i \bar U_j H_u$ then carries a $X$-charge $n_{ij}$\footnote{For
convenience, we will assume that $n_{ij} \geq 0$.}, which is simply the
sum of the $X$-charges of the fields entering the coupling: $n_{ij} =
X_{Q_i} + X_{\bar U_i} + X_{H_u}$ ($i$ and $j$ are generation indices,
$Q_i$ is the quark doublet of the $i^{th}$ generation, $\bar U_j$ the
quark singlet of the $j^{th}$ generation, and $H_u$ the Higgs doublet
that gives a mass to the up quarks). If $n_{ij} \neq 0$, the coupling is
forbidden by $U(1)_X$, and the corresponding mass matrix entry is zero.
When $n_{33} = 0$ and $n_{ij} \neq 0$ for $(i,j) \neq (3,3)$, only the
top quark coupling is allowed, and the up quark mass matrix has the form
(\ref{eq:hierarchy}). The other Yukawa couplings are then generated from
non-renormalizable interactions involving a singlet field $\theta$ with
$X$-charge $X_{\theta} = - 1$:
\begin{displaymath}
  Q_i \bar U_j H_u \left( {\theta \over M} \right)^{n_{ij}}
\end{displaymath}
where $M$ is a large scale characteristic of the underlying theory
(typically $M \sim M_{Planck}$ or $M_{GUT}$). When the MSSM singlet
$\theta$ acquires a vacuum expectation value, which breaks spontaneously
$U(1)_X$, effective Yukawa couplings are generated:
\begin{equation}
  Y_{ij} \sim \left( \frac{<\theta>}{M} \right) ^{n_{ij}}
\end{equation}
with their orders of magnitude fixed by their charges under $U(1)_X$. It
follows that the structure of the mass matrix is determined by the
family symmetry (we use the notation $\epsilon = <\theta> / M$):
\begin{equation}
  \frac{{\cal M}_u}{m_t}  \sim  \left( \begin{array}{ccc}
    \epsilon^{\: n_{11}} & \epsilon^{\: n_{12}} &
	\epsilon^{\: n_{13}} \\
    \epsilon^{\: n_{21}} & \epsilon^{\: n_{22}} &
	\epsilon^{\: n_{23}} \\
    \epsilon^{\: n_{31}} & \epsilon^{\: n_{32}} & 1
	\end{array} \right)
\end{equation}
where only the order of magnitude of each entry is given. Since $U(1)_X$
is broken below the scale $M$, $\epsilon$ is a small parameter
(typically $\epsilon \sim 0.1$), and a hierarchy between Yukawa
couplings naturally appears. The diagonalization of ${\cal M}_u$ and
${\cal M}_d$ (which is obtained in the same way than ${\cal M}_u$) then
relates the mass and mixing hierarchy of the quarks to their charges
under $U(1)_X$.

\subsection{Lepton sector}

 \indent
If the lepton fields $L_i$, $\bar E_i$ and $\bar N_i$ carry $X$-charge,
the previous mechanism also works for the lepton mass matrices ${\cal
M}_e$, ${\cal M}_D$ and ${\cal M}_M$. In particular, the structure of
the Dirac and the Majorana matrices is determined by the family symmetry
{\it without any anstaz}.

More precisely, ${\cal M}_D$ and ${\cal M}_M$ are generated from higher
order operators of the type\footnote{For $p_{ij} \geq 0$ and $q_{ij}
\geq 0$.}
\begin{eqnarray}
  L_i \bar N_j H_u \left( \frac{\theta}{M} \right) ^{p_{ij}}  &  &
	p_{ij} = X_{L_i} + X_{\bar N_j} + X_{H_u}
\end{eqnarray}
for the Dirac matrix and
\begin{eqnarray}
  M \bar N_i \bar N_j \left( \frac{\theta}{M} \right) ^{q_{ij}}  &  &
	q_{ij} = X_{\bar N_i} + X_{\bar N_j}
\end{eqnarray}
for the Majorana matrix. After breaking of $U(1)_X$, one obtains:
\begin{eqnarray}
  ({\cal M}_D)_{ij} \sim m \left( \frac{<\theta>}{M} \right) ^{\:p_{ij}}
	&  &  
  ({\cal M}_M)_{ij} \sim M \left( \frac{<\theta>}{M} \right) ^{\:q_{ij}}
\end{eqnarray}
The structures of the Dirac and the Majorana matrices, and consequently
the neutrino masses and mixing angles, are determined by the neutrino
charges under $U(1)_X$. No ansatz is required.

Several groups have studied neutrino mass models based on a family
symmetry \cite{DLLRS,GN,BLR}. In the following, we present two of them.

%%%%%%%%%%%%%%%%%%%%%%%%%%%%%%%%%%%%%%%%%%%%%%%%%%%%%%%%%%%%%%%%%%%%%%%%%

\section{Examples of neutrino mass models with a U(1) family symmetry}

\subsection{Model 1}

 \indent
The assumptions of this model \cite{BLR} are the following: (a) the
anomalies of the horizontal $U(1)_X$ are compensated for by an
appropriate mechanism (Green-Schwarz); (b) the dominant entry in each
mass matrix is the (3,3) entry; (c) the $X$-charges of all mass terms
are positive ($p_{ij} \geq 0$, $q_{ij} \geq 0$). (b) and (c) allow to
make a simple analysis. Note that ${\cal M}_D$ is not assumed to be
symmetric, which gives us more liberty in the choice of the parameters
of the $X$-charge.

With these assumptions, the light neutrino masses are:
\begin{eqnarray}
m_{\nu_e} & \sim & {m^2_3 \over M_3} \;
	\epsilon^{\: 2(X_{L_1} - X_{L_3})}  \nonumber \\
m_{\nu_\mu} & \sim & {m^2_3 \over M_3} \;
	\epsilon^{\: 2(X_{L_2} - X_{L_3})} 
	\label{eq:light} \\
m_{\nu_\tau} & \sim & {m^2_3 \over M_3}  \nonumber
\end{eqnarray}
The $\nu_\tau$ mass is given by the usual seesaw formula ($M_3$ is the
mass of the heaviest $RH$ neutrino, $m_3$ the largest Dirac mass),
whereas the other neutrino masses are suppressed relative to
$m_{\nu_{\tau}}$ by powers of the small breaking parameter $\epsilon$.
Note that the hierarchy depends only on the $X$-charges of the lepton
doublets $L_i$.
The lepton mixing matrix is:
\begin{equation} 
V_L \sim R_{\nu} \sim  \left( \begin{array}{ccc}
1 & \epsilon^{\: |X_{L_1} - X_{L_2}|} & 
  \epsilon^{\: |X_{L_1} - X_{L_3}|}  \\
\epsilon^{\: |X_{L_1} - X_{L_2}|} & 1 & 
  \epsilon^{\: |X_{L_2} - X_{L_3}|}  \\
\epsilon^{\: |X_{L_1} - X_{L_3}|} & 
  \epsilon^{\: |X_{L_2} - X_{L_3}|} & 1  \end{array} \right)
\end{equation}
$V_L$ has the same structure than $R_{\nu}$ because, due to the
assumptions of the model, the diagonalizing matrices for the neutrino
masses ($R_{\nu}$) and the charged lepton masses ($R_{e}$) have the same
structure.

Note that the light neutrino spectrum would not have been affected if,
instead of adding one $RH$ neutrino per family, we had introduced an
arbitrary number of such heavy fields.

This model has several remarkable features. First, it is worth noting
that the neutrino mass and mixing hierarchies do {\it not} depend on the
particular form of the Majorana matrix. This is a great difference with
most seesaw models. The reason for this is that the dependences of
${\cal M}_D$ and ${\cal M}_M$ on the heavy neutrino charges compensate
for each other in the matrix ${\cal M}_{\nu}$. Secondly, the mass
spectrum obtained is naturally hierarchical\footnote{Mass degeneracies
are not excluded, but in this case the model is less predictive, since
the mass difference between almost degenerate neutrinos cannot be
related to the lepton $X$-charges.}, without hierarchy inversion:
$m_{\nu_e} \ll m_{\nu_\mu} \ll m_{\nu_\tau}$. Finally, the mixing angles
and the mass ratios are usually related by:
\begin{equation}
  \sin^2 \theta_{ij} \sim {m_{\nu_i} \over m_{\nu_j}}
\label{eq:angle}
\end{equation}
These relations, which generalize (\ref{eq:angle1}), are common to
numerous seesaw models. They imply $V_{e \nu_\mu} V_{\mu \nu_\tau} \sim
V_{e \nu_\tau}$ in lepton charged current, in analogy with $V_{us}
V_{cb} \sim V_{ub}$ in quark charged current.

The experimental data on solar neutrinos and atmospheric neutrinos put
constraints on the parameters of the model. For example, if one wants to
explain simultaneously the solar neutrino data by MSW $\nu_e \rightarrow
\nu_\mu$ transitions, and the atmospheric neutrino data by $\nu_\mu -
\nu_\tau$ oscillations, one must choose\footnote{We only address the
case of a hierarchical mass spectrum: $m_{\nu_e} \ll m_{\nu_\mu} \ll
m_{\nu_\tau}$.}:
\begin{eqnarray}
  X_{L_1} - X_{L_3} = 3  &  &  X_{L_2} - X_{L_3} = 1
\end{eqnarray}
which leads to the following spectrum:
\begin{eqnarray}
  \left\{ \begin{array}{ll}
	m_{\nu_e} \sim 10^{-5} & eV  \\
	m_{\nu_\mu} \sim 5.10^{-3} & eV  \\  
	m_{\nu_\tau} \sim 0.1 & eV
  \end{array} \right.  & &  \left\{ \begin{array}{l}
	sin^2 \: 2 \theta_{e \mu} \simeq 2.10^{-3} - 4.10^{-2}  \\
	sin^2 \: 2 \theta_{\mu \tau} \simeq 5.10^{-2} - 0.8
  \end{array} \right.
\end{eqnarray}
The uncertainties in the mixing angles are due to the fact that the mass
matrix entries are determined by the family symmetry up to a factor of
order one. Note that, in order to avoid fine-tuning, we have only
addressed the hierarchical case, which makes it quite difficult to
obtain a large mixing angle [see (\ref{eq:angle})], as required by the
atmospheric neutrino data. Furthermore, the tau neutrino is too light to
be an interesting candidate for dark matter. However, if one ignores the
atmospheric neutrino problem, it is possible to obtain a relevant
$\nu_\tau$ for cosmology and to explain the solar neutrino data at once.

\subsection{Model 2}

 \indent
The assumptions of this model \cite{DLLRS}, which was proposed first,
are quite different from the previous one: (a) all mass matrices are
symmetric\footnote{This arises naturally in left-right symmetric GUT's,
like $SU(3)^3$ or $E_6$.}, which reduces the number of independent
parameters; (b) the $X$-charges of the mass terms can be negative, and
the existence of a pair $(\theta, \bar \theta)$ of singlets with
opposite $X$-charges is assumed; (c) the heavy neutrino Majorana masses
are generated from the coupling to a singlet Higgs boson $\Sigma$ with
charge $X_{\Sigma}$ [$<\Sigma> \bar N_i \bar N_j$], which gives rise to
a discrete spectrum of possible Majorana matrices, depending on
$X_\Sigma$. With (a) and (b), the Dirac matrix takes the form:
\begin{equation}
  {\cal M}_D \sim \left( \begin{array}{ccc}
    \epsilon^{\: |p_{11}|} & \epsilon^{\: |p_{12}|} &
	\epsilon^{\: |p_{13}|} \\
    \epsilon^{\: |p_{12}|} & \epsilon^{\: |p_{22}|} &
	\epsilon^{\: |p_{23}|} \\
    \epsilon^{\: |p_{13}|} & \epsilon^{\: |p_{23}|} & 1
	\end{array}  \right)
\end{equation}
(c) implies that, for a given ${\cal M}_D$ (corresponding to a given
assignment of lepton $X$-charges), eg
\begin{equation}
  {\cal M}_D \sim \left( \begin{array}{lll}
    \epsilon^{\: 16} & \epsilon^{\: 6} & \epsilon^{\: 8} \\
    \epsilon^{\: 6} & \epsilon^{\: 4} & \epsilon^{\: 2} \\
    \epsilon^{\: 8} & \epsilon^{\: 2} & 1   \end{array}  \right)
\end{equation}
several ${\cal M}_M$ are possible, depending on $X_\Sigma$. As a
consequence, the light neutrino spectrum depends on the value of
$X_\Sigma$. This is illustrated by the following two examples:
\begin{eqnarray}
    X_\Sigma = X_{H_u} \; \; \; \; \; \;
  &
    {\cal M}^{diag}_\nu \sim \left( \begin{array}{lll}
    \epsilon^{\: 11} & 0 & 0 \\
    0 & \epsilon^{\: 7} & 0 \\
    0 & 0 & 1   \end{array}  \right)
  &
    R_\nu \sim \left( \begin{array}{lll}
    1 & \epsilon^{\: 3} & \epsilon^{\: 5} \\
    \epsilon^{\: 3} & 1 & \epsilon^{\: 2} \\
    \epsilon^{\: 5} & \epsilon^{\: 2} & 1   \end{array}  \right)
\end{eqnarray}
\begin{eqnarray}
    X_\Sigma = {3 \over 2} X_{H_u}
  &
    {\cal M}^{diag}_\nu \sim \left( \begin{array}{lll}
    \epsilon^{\: 11} & 0 & 0 \\
    0 & \epsilon^{\: 5} & 0 \\
    0 & 0 & \epsilon^{\: -1}   \end{array}  \right)
  &
    R_\nu \sim \left( \begin{array}{lll}
    1 & \epsilon^{\: 3} & \epsilon^{\: 7} \\
    \epsilon^{\: 3} & 1 & \epsilon^{\: 2} \\
    \epsilon^{\: 7} & \epsilon^{\: 2} & 1   \end{array}  \right)
\end{eqnarray}

Thus, in this model, the neutrino mass ratios and mixing angles are
determined by the lepton $X$-charges up to a discrete
ambiguity\footnote{However, if left-right symmetry is assumed, the value
of $X_\Sigma$ is fixed by anomaly considerations.}. The mass spectrum
thus obtained is always hierarchical, and mass degeneracy requires
fine-tuning. Note that, due to assumption (b), the mass-angle relation
(\ref{eq:angle}) is not automatically satisfied. Furthermore, the
matrices that diagonalize the charged lepton and neutrino mass matrices
do not have the same structure, hence the mixing angles relevant for
neutrino oscillations are not simply given by $R_\nu$.

Finally, the model is able to reproduce the experimental neutrino data.
$m_{\nu_\tau}$ is in the interesting range for cosmology ($1-10 \; GeV$)
for reasonable values of $<\Sigma>$. The small angle MSW solution to the
solar neutrino problem can be accommodated. However, a large mixing
angle, as required by the atmospheric neutrino data, cannot be obtained
without fine-tuning.

\section{Conclusion}

 \indent
We have presented two neutrino mass models based on an abelian family
symmetry. The great advantage of such models is that the structure of
the Dirac and the Majorana matrices is entirely determined by the
symmetry, therefore the neutrino masses and mixing angles are predicted
without any ansatz. Furthermore, the fact that the same symmetry is able
to explain the observed fermion mass hierarchy and simultaneously
constrains the neutrino spectrum sets an interesting connection between
two fundamental problems in particle physics. Unfortunately, the lepton
$X$-charges are constrained, but not fully determined by the model,
which reduces its predictive power. Moreover, it is difficult to
understand mass degeneracies as well as mixing angles of order one,
which are necessary to account for all neutrino data.

The main part of this talk is based on work done in collaboration with
P. Bin\'etruy and P. Ramond.

%%%%%%%%%%%%%%%%%%%%%%%%%%%%%%%%%%%%%%%%%%%%%%%%%%%%%%%%%%%%%

\end{document}